# BIOMERO 2.0: end-to-end FAIR infrastructure for bioimaging data import, analysis, and provenance

Torec T. Luik[1*], Joost de Folter [1,2], Rodrigo Rosas-Bertolini[3], Eric A.J. Reits[1], Ron A. Hoebe[1¶], Przemek M. Krawczyk[1,4¶*]

[1] Amsterdam UMC, Department of Medical Biology, Amsterdam, North-Holland, 1105AZ, The Netherlands

[2] The Francis Crick Institute, London NW1 1BF, United Kingdom

[3] Independent Researcher, Brussels, Belgium

[4] lead contact

* Corresponding authors

E-mail: p.krawczyk@amsterdamumc.nl (PK); t.t.luik@amsterdamumc.nl (TL)

¶ These authors contributed equally to this work.

## Summary

We present BIOMERO 2.0, a major evolution of the BIOMERO framework that transforms OMERO into a FAIR-compliant (findable, accessible, interoperable, and reusable), provenance-aware bioimaging platform. BIOMERO 2.0 integrates data import, preprocessing, analysis, and workflow monitoring through an OMERO.web plugin and containerized components. The importer subsystem facilitates in-place import using containerized preprocessing and metadata enrichment via forms, while the analyzer subsystem coordinates and tracks containerized analyses on high-performance computing systems via the BIOMERO Python library. All imports and analyses are recorded with parameters, versions, and results, ensuring real-time provenance accessible through integrated dashboards. This dual approach places OMERO at the heart of the bioimaging analysis process: the importer ensures provenance from image acquisition through preprocessing and import into OMERO, while the analyzer records it for downstream processing. These integrated layers enhance OMERO's FAIRification, supporting traceable, reusable workflows for image analysis that bridge the gap between data import, analysis, and sharing.

## Introduction

Modern bioimaging experiments, especially high-content and high-throughput microscopy, generate millions of images along with complex experimental metadata[1]. Managing such data effectively requires integrated platforms that support FAIR data principles, including detailed provenance tracking[1–3]. OMERO (Open Microscopy Environment Remote Objects)[4] is a leading open-source repository for biological images and associated metadata. It provides a centralized database for storing images, metadata, and annotations, with tools for collaborative sharing and analysis[1]. However, standard OMERO workflows typically rely on external import clients (e.g. the

OMERO.insight desktop importer) and separate analysis pipelines, which complicates end-to-end traceability of data and processing steps[1,5]. For true data reuse and reproducibility (FAIR R1.2), the full history of how data were generated and processed must be captured in machine-readable form, including who created them, how they were transformed, and by which workflows[2].

To address these challenges, we developed BIOMERO (Bioimage analysis in OMERO)[6]. BIOMERO acts as a bridge between OMERO and high-performance computing (HPC) clusters, enabling researchers to launch FAIR-compliant image analysis workflows directly from the OMERO interface[6]. BIOMERO simplifies the process: users select an analysis script in OMERO.web, BIOMERO transfers the data to the compute cluster, runs containerized analyses under a job scheduler, and returns results to OMERO[6]. This approach brought automated, scalable processing into OMERO without requiring users to master HPC or workflow management. While BIOMERO streamlined analysis, gaps remained in data ingest and metadata capture.

In this work we extend BIOMERO to make data ingestion and provenance capture seamless parts of the OMERO ecosystem. We developed a web-based importer that integrates complex datasets and their associated metadata directly into OMERO. By enabling in-place imports via OMERO's web interface, this method eliminates the need for external desktop tools and reduces data duplication by keeping the raw data in its original storage location. This method simplifies data ingestion, ensures data integrity, and keeps data management efficient. Additionally, we enhanced the OMERO web client with an updated OMERO.forms plugin for structured metadata entry, allowing users to define customized, JSON-schema–based forms for their experiments. This plugin records form and input versions immutably[7,8], ensuring that users can always review exactly what metadata were entered and which version was used, thus providing built-in provenance. BIOMERO 2.0 also includes detailed tracking of data processing steps through a backend database and web dashboard, accessible via the "OMERO.biomero" plugin in OMERO web UI[9]. Each analysis run is uniquely identified and linked to its inputs and outputs, fully capturing the processing history of the data.

In summary, BIOMERO 2.0 unifies data import, structured metadata annotation, and analysis in a single, FAIR-compliant environment. By combining an in-place importer, enhanced user interface, and immutable provenance capture, BIOMERO ensures that every image and its analysis history are traceable from acquisition through processing[2,8]. This system significantly improves the reusability and reproducibility of bioimaging data, filling gaps in both OMERO and BIOMERO v1 systems. By enabling FAIR-compliant workflows, it offers a level of data traceability and transparency that was previously not possible with open-source alternatives.

# Results

## Architecture and design

BIOMERO 2.0 was developed to address the need for a fully integrated, FAIR-compliant bioimaging infrastructure, focusing on traceability, reproducibility, and efficient data management throughout the research pipeline. The system is designed to tackle the challenges of data ingestion, metadata annotation, analysis execution, and provenance tracking within OMERO, while ensuring that bioimaging workflows are transparent and reproducible. It combines these functionalities into modular, interoperable components, ensuring FAIR compliance and enabling access via both OMERO.web and programmatic API calls.

To achieve these goals, BIOMERO 2.0 is structured around several architectural principles: (1) modularity, where each component (ingestion, analysis, provenance tracking) is independent but integrates smoothly into the overall system; (2) scalability, which allows the system to manage large-scale, high-throughput datasets; (3) interoperability, enabling the system to integrate with different platforms and environments; and (4) user accessibility, ensuring that the system is intuitive for both technical and non-technical users via OMERO.web and API endpoints.

Figure 1 provides a schematic overview of the BIOMERO 2.0 architecture. At its core is the original BIOMERO 1.0 setup, which was based on the BIOMERO Python library (labeled BIOMERO.analyzer, in blue). The system extends this setup with the OMERO.biomero web interface plugin, which offers access to both the analyzer and the new importer. BIOMERO.scripts continue to run behind the scenes for analysis, now accessed via a reactive and scalable script interface. The OMERO.biomero plugin also introduces the first web-based interface for direct in-place imports from remote storage, which eliminates the need for external tools and ensures data integrity by keeping raw data in its original storage location. The following sections will delve into the details of these subsystems.

## The BIOMERO.importer enables automated preprocessing and in-place data import

Figure 2 provides a schematic overview of the importer architecture, while Figure S1 presents a detailed flowchart of the data flow. The importer operates on a designated data folder, which can be remote storage, a mounted disk, or a network-accessible location, and requires read and write permissions. It continuously monitors the database for pending import orders and executes them in two stages: first, an optional preprocessing step, and second, an in-place import into OMERO. This design allows the system to handle incoming data automatically while keeping raw files on the original storage.

The preprocessing step allows the importer to handle data formats unsupported by OMERO or BioFormats by running containerized workflows that convert them into compatible representations, such as OME-TIFF, OME-ZARR, or just to create pyramidal images.

BIOMERO.importer executes these conversions using Podman containers, writing the processed data back to storage before in-place import. All operations are tracked automatically in the back end without user involvement. Example converters for Pico database and Leica imaging data are described in the Note S1 and documented on GitHub.

BIOMERO.importer's key feature is its ability to perform in-place imports from attached storage, initiated through the OMERO.biomero web interface. It extends OMERO.cli capabilities, previously restricted to users with direct server access. Unlike the OMERO.insight desktop client, which only supports uploads and lacks web integration, BIOMERO.importer keeps raw data on the remote storage after import instead of duplicating it on the OMERO.server. This approach aligns with BIOMERO 1.0's design philosophy of offloading computation to external HPC resources.

BIOMERO.importer also enables end-users to perform in-place imports of OME-ZARR files via the OMERO.biomero web interface. This functionality requires the OMERO.server to be upgraded with Glencoe's OMERO Zarr Pixel Buffer[10], available under a GPL-2.0 license. Although OMERO.cli provides the underlying functionality, direct access is typically restricted to server-level users, and neither OMERO.insight nor the web interface provide this capability for end-users. This makes BIOMERO.importer a critical component for integrating OME-ZARR data into OMERO workflows.

## User Interface enables real accessibility

Accessibility extends beyond data retrieval; it means making interaction intuitive and available to all users. To achieve this, OMERO.biomero provides full web-based access to both data import and analysis, supported by complementary visualization tools. Instead of relying on command-line or API access, users can now monitor, explore, and manage their workflows entirely through the browser.

A key enhancement is the integration of a Metabase service that generates interactive dashboards directly from our database tables. These dashboards offer live system overviews, automatically updating as background processes run. Embedded within the OMERO.biomero plugin via iFrames, they allow users to track progress and view analytics without leaving the interface. Dashboards are fully customizable and can summarize workflow usage, system activity, and other statistics. Together, the PostgreSQL database and Metabase dashboards form the BIOMERO.db system (Figures 1–2). Figures S6, S15 and S16 show screenshots of the Metabase dashboard that is integrated into the OMERO.biomero user interface.

Figures S2-S21 show the new OMERO.biomero UI. The importer module includes four tabs: Import Images, Import Screens, Monitor, and Admin. In the import tabs, users first select a target location in OMERO (a dataset or screen) and then choose source data from remote storage. Selected items are added to an import list and queued for processing. Both import tabs also integrate OMERO.forms for entering experimental metadata, even while imports are running in the background; this is shown in Figures S3 and S4. When a new import is started, the web interface creates an order in the BIOMERO.db, which the BIOMERO.importer service then executes asynchronously, including any assigned preprocessing. Users can close their browser or continue working in OMERO while imports proceed, with live tracking available in the Monitor tab. The Admin tab, visible only to administrators, allows mapping of OMERO groups to subfolders on the remote storage, restricting access to relevant data areas.

The analyzer module offers three tabs: Run, Status, and Admin. The Run tab provides workflow-first access to all BIOMERO workflows. Users browse workflows through searchable cards showing titles, descriptions (e.g., segmentation), and links to GitHub and DockerHub. After selecting a workflow, users choose input data, adjust the workflow parameters, and define output handling options. This design is far more scalable and user-friendly than the previous OMERO.scripts interface. The Status tab includes a Metabase dashboard displaying real-time progress of workflows on HPC, including queue or runtime states. This is enabled by an event-sourcing subsystem added on top of BIOMERO 1.0, which logs all actions to the new BIOMERO.db. From this log, views can be generated for progress monitoring or administrative reporting, such as linking users to specific HPC jobs for accounting. Finally, the Admin tab (admin-only) provides web-based configuration of BIOMERO.analyzer settings, including workflow registration and Slurm parameters, eliminating the need for direct server access as in version 1.0.

## Provenance across all subsystems for end-to-end coverage

To support reuse in line with FAIR principle R1.2[2], BIOMERO 2.0 captures the full history of each dataset: its origin, who created it, which workflows were applied, and any preprocessing or transformations. This end-to-end provenance is achieved through three complementary layers:

- OMERO.forms**:** Provides structured, shareable metadata forms for manual entry, enabling core facilities and the community to define the experimental metadata that must be captured. Forms can implement REMBI[11] submodules or other community standards[12], support multiple experimental purposes, and track changes with versioning. The latest values are stored directly on the image object, ensuring consistent, machine-readable metadata from the start of data acquisition. This layer focuses on pre-acquisition experimental details and promotes best practices without imposing undue burden on users.
- BIOMERO.importer: Records metadata about the original file, the import order, the responsible user, and any preprocessing applied via versioned containers. It can also parse output metadata from preprocessing containers or CSV files stored alongside the raw data into OMERO key-value pairs. Each import is assigned a unique ID, allowing users to retrieve all images associated with a given order once importing is completed. This ensures traceability of all data entering the system.

- BIOMERO.analyzer: Logs all workflow actions in an eventsourcing database. From these logs, both real-time progress dashboards and complete provenance records can be generated. Each workflow instance records all inputs, parameters, versions, and outputs, linking derived images (e.g., masks or segmentations) back to the workflow that created them. This allows users to reconstruct the full history of analysis, even for jobs executed asynchronously on HPC resources.

By integrating these three layers, BIOMERO 2.0 embeds metadata throughout the data lifecycle, from experimental design and acquisition through import and preprocessing to analysis, ensuring autonomous, consistent provenance capture. All metadata fields are indexed and fully searchable within OMERO, reducing errors, saving user time, and enabling reproducible, FAIR-compliant microscopy research.

## Discussion

BIOMERO 2.0 significantly advances FAIR bioimaging workflows by combining data management (DM) and data analysis (DA) within a single, integrated platform. The modular design of BIOMERO 2.0 facilitates interoperability with external containers for both image analysis and preprocessing. This modularity enables the reuse of containers in different systems, environments, or labs, providing flexibility and scalability. At the same time, all actions, including container versions, parameters, and execution details, are meticulously tracked for provenance, ensuring transparency and reproducibility. This approach enhances workflow efficiency while maintaining data integrity and traceability across the entire research pipeline.

The integration of provenance into every stage of the workflow is central to BIOMERO 2.0. Rather than operating independently, provenance is embedded directly into the data management and analysis infrastructure. With OMERO.biomero, users can track the precise containers and versions used during image analysis and preprocessing, and this information is captured in the system's metadata. This ensures complete data traceability and facilitates reproducibility, even in complex, multi-step workflows.

While BIOMERO 2.0 automates much of the process for FAIR compliance, manual metadata entry remains a vital component. Through OMERO.forms, administrators can create structured templates for users, guiding them in entering the correct experimental metadata. This guided input ensures that users provide essential information without errors, while BIOMERO automatically tracks the provenance of analysis tasks and job executions. This combination of user input and automated provenance tracking makes it easier to maintain metadata consistency and supports the overall FAIR objectives.

Looking ahead, BIOMERO 2.0 is positioned to integrate AI tools that support both metadata entry and workflow design. Large language models (LLMs) can assist in converting user descriptions into standardized, precise metadata, ensuring that all data is properly annotated in line with FAIR principles. LLMs could also help guide users in selecting appropriate analysis tasks or designing workflows based on the specific needs of their experiment. While AI may not yet manage provenance in full, BIOMERO ensures that all actions taken through the platform are fully captured and traceable, supporting data reproducibility and transparency at each stage. With this foundation, AI can play a critical role in enhancing both metadata quality and the sophistication of bioimaging data workflows.

In conclusion, BIOMERO 2.0 goes beyond merely aligning with FAIR principles—it actively enables their implementation. By integrating data management, image analysis, and provenance tracking into one unified platform, BIOMERO provides essential infrastructure for reproducible bioimaging research. Future developments, including AI-assisted metadata compliance and more advanced support for standardized, shareable workflows, will strengthen BIOMERO's ability to support the bioimaging community's transition to fully transparent, traceable, and FAIR-compliant research.

# Experimental Procedures

## Resource availability

**Lead contact**

Further information and requests for resources should be directed to and will be fulfilled by the lead contact, Przemek M. Krawczyk (p.krawczyk@amsterdamumc.nl).

**Materials availability**

This study did not generate new unique reagents.

**Data and code availability**

- All original code for the BIOMERO.analyzer, consisting of the BIOMERO Python library and BIOMERO.scripts, is publicly available as of the date of publication at GitHub (https://github.com/NL-BioImaging/biomero and https://github.com/NL-BioImaging/biomero-scripts ) and PyPI (https://pypi.org/project/biomero ). BIOMERO Python library is available under the permissive Apache 2.0 (https://www.apache.org/licenses/LICENSE-2.0 ) license, while BIOMERO.scripts is available under the copyleft GPL 2.0 (https://www.gnu.org/licenses/old-licenses/gpl-2.0.en.html ) license. All original code for the BIOMERO.importer is publicly available as of the date of publication at GitHub (https://github.com/Cellular-Imaging-Amsterdam-UMC/BIOMERO.importer ) and PyPI (https://pypi.org/project/biomero-importer/ ), under the permissive Apache 2.0 (https://www.apache.org/licenses/LICENSE-2.0 ) license.

- All original code for the OMERO.forms plugin is publicly available as of the date of publication at GitHub (https://github.com/NL-BioImaging/OMERO.forms ) and PyPI (https://pypi.org/project/omero-forms/ ), under the copyleft AGPL 3.0 (https://www.gnu.org/licenses/agpl-3.0.en.html/ ) license.
- All original code for preprocessing containers is publicly available as of the date of publication at GitHub (https://github.com/Cellular-Imaging-Amsterdam-UMC/ConvertLeica-Docker and https://github.com/NL-BioImaging/biomero-converter ), under the permissive Apache 2.0 (https://www.apache.org/licenses/LICENSE-2.0 ) license.
- All original code for NL-BIOMERO (including deployment scenarios and Dockerfiles) is publicly available as of the date of publication at GitHub (https://github.com/Cellular-Imaging-Amsterdam-UMC/NL-BIOMERO ), under the permissive BSD 2-Clause (https://opensource.org/license/bsd-2-clause ) license. Docker images are publicly available, versioned with every release, for all our services on DockerHub (https://hub.docker.com/u/cellularimagingcf ).

# Acknowledgments

This publication is part of the project NL-BioImaging-AM (with project number 184.036.012 of the National Roadmap research programme which is (partly) financed by the Dutch Research Council (NWO). Special thanks to our collaborators within the NL-BioImaging project for their valuable contributions. We also appreciate the support and insights provided by the bioimaging community and OME team on image.sc, which greatly enhanced the quality and scope of this research. Icons used in images are (adapted) from Material Design Icons (https://pictogrammers.com/library/mdi/ ) and OpenClipart (https://openclipart.org).

# Author Contributions

Conceptualization, TL and RR and ER and RH and PK; Methodology, TL and PK and RH and JF; Software, TL and RR and JF and PK; Validation, TL and RH and JF; Writing – Original draft, TL; Writing – Review & Editing, PK and RH and ER and RR and JF and TL; Funding Acquisition, ER; Supervision, PK and RH

# Declaration of Interests

The authors declare no competing interests.

# Declaration of generative AI and AI-assisted technologies in the writing process

During the preparation of this work the author(s) used ChatGPT in order to improve readability and language of the text and generate drafts for paragraphs. After using this tool/service, the author(s) reviewed and edited the content as needed and take(s) full responsibility for the content of the publication.

# References

1. Massei, R., Busch, W., Serrano-Solano, B., Bernt, M., Scholz, S., Nicolay, E.K., Bohring, H., and Bumberger, J. (2025). High-content screening (HCS) workflows for FAIR image data management with OMERO. Sci Rep *15*, 16236. https://doi.org/10.1038/s41598-025-00720-0.

2. R1.2: (Meta)data are associated with detailed provenance GO FAIR. https://www.go-fair.org/fair-principles/r1-2-metadata-associated-detailed-provenance/.

3. Wilkinson, M.D., Dumontier, M., Aalbersberg, Ij.J., Appleton, G., Axton, M., Baak, A., Blomberg, N., Boiten, J.-W., da Silva Santos, L.B., Bourne, P.E., et al. (2016). The FAIR Guiding Principles for scientific data management and stewardship. Sci Data *3*, 160018. https://doi.org/10.1038/sdata.2016.18.

4. Allan, C., Burel, J.-M., Moore, J., Blackburn, C., Linkert, M., Loynton, S., MacDonald, D., Moore, W.J., Neves, C., Patterson, A., et al. (2012). OMERO: flexible, model-driven data management for experimental biology. Nat Methods *9*, 245–253. https://doi.org/10.1038/nmeth.1896.

5. Import data using the Desktop Client — OMERO guide latest documentation https://omero-guides.readthedocs.io/en/latest/upload/docs/import-desktop-client.html.

6. Luik, T.T., Rosas-Bertolini, R., Reits, E.A.J., Hoebe, R.A., and Krawczyk, P.M. (2024). BIOMERO: A scalable and extensible image analysis framework. Patterns *5*. https://doi.org/10.1016/j.patter.2024.101024.

7. NL-BioImaging/OMERO.forms (2025). (NL-BioImaging AM).

8. Russell, D.P.W., and Sorger, P.K. (2017). Maintaining the provenance of microscopy metadata using OMERO.forms software. Preprint at bioRxiv, https://doi.org/10.1101/109199 https://doi.org/10.1101/109199.

9. Cellular-Imaging-Amsterdam-UMC/OMERO.biomero (2025). (Cellular Imaging, Amsterdam UMC).

10. glencoesoftware/omero-zarr-pixel-buffer (2025). (Glencoe Software).

11. Sarkans, U., Chiu, W., Collinson, L., Darrow, M.C., Ellenberg, J., Grunwald, D., Hériché, J.-K., Iudin, A., Martins, G.G., Meehan, T., et al. (2021). REMBI: Recommended Metadata for Biological Images—enabling reuse of microscopy data in biology. Nat Methods *18*, 1418–1422. https://doi.org/10.1038/s41592-021-01166-8.

12. Hosseini, R., Vlasveld, M., Willemse, J., van de Water, B., Le Dévédec, S.E., and Wolstencroft, K.J. (2023). FAIR High Content Screening in Bioimaging. Sci Data *10*, 462. https://doi.org/10.1038/s41597-023-02367-w.

# Figure Legends

Figure 1 – Schematic overview of the BIOMERO 2.0 architecture. BIOMERO 2.0 extends the original BIOMERO 1.0 framework (blue) with new components (orange) to enable full provenance capture across imaging workflows. The OMERO.biomero web interface now serves as the central user entry point, providing access to both image analysis (blue, left) and data import (orange, right) directly from the web. Through OMERO.forms, it promotes best practices for structured metadata capture. While BIOMERO 1.0 remains accessible via BIOMERO.scripts, the new web interface adds a responsive layer for improved usability. The BIOMERO.scripts and BIOMERO.analyzer library now record all parameters and actions to both the BIOMERO.db and OMERO's metadata system. The new BIOMERO.importer supports in-place data import from remote storage, with optional preprocessing steps such as format conversion. All components integrate through the BIOMERO.db, which also powers user-facing dashboards summarizing import and analysis activities.

Figure 2 – The BIOMERO.importer enables automated preprocessing and in-place data import. The BIOMERO.importer service automates the ingestion of imaging data into OMERO while maintaining full provenance. Users initiate an import via the OMERO.biomero interface by selecting source data (image or screen) and a destination. This action creates an order in the BIOMERO.db, which is continuously monitored by the importer. When a new order is detected, the importer optionally executes a defined preprocessing container on the selected data before performing an in-place import into OMERO, under the user's account and group.

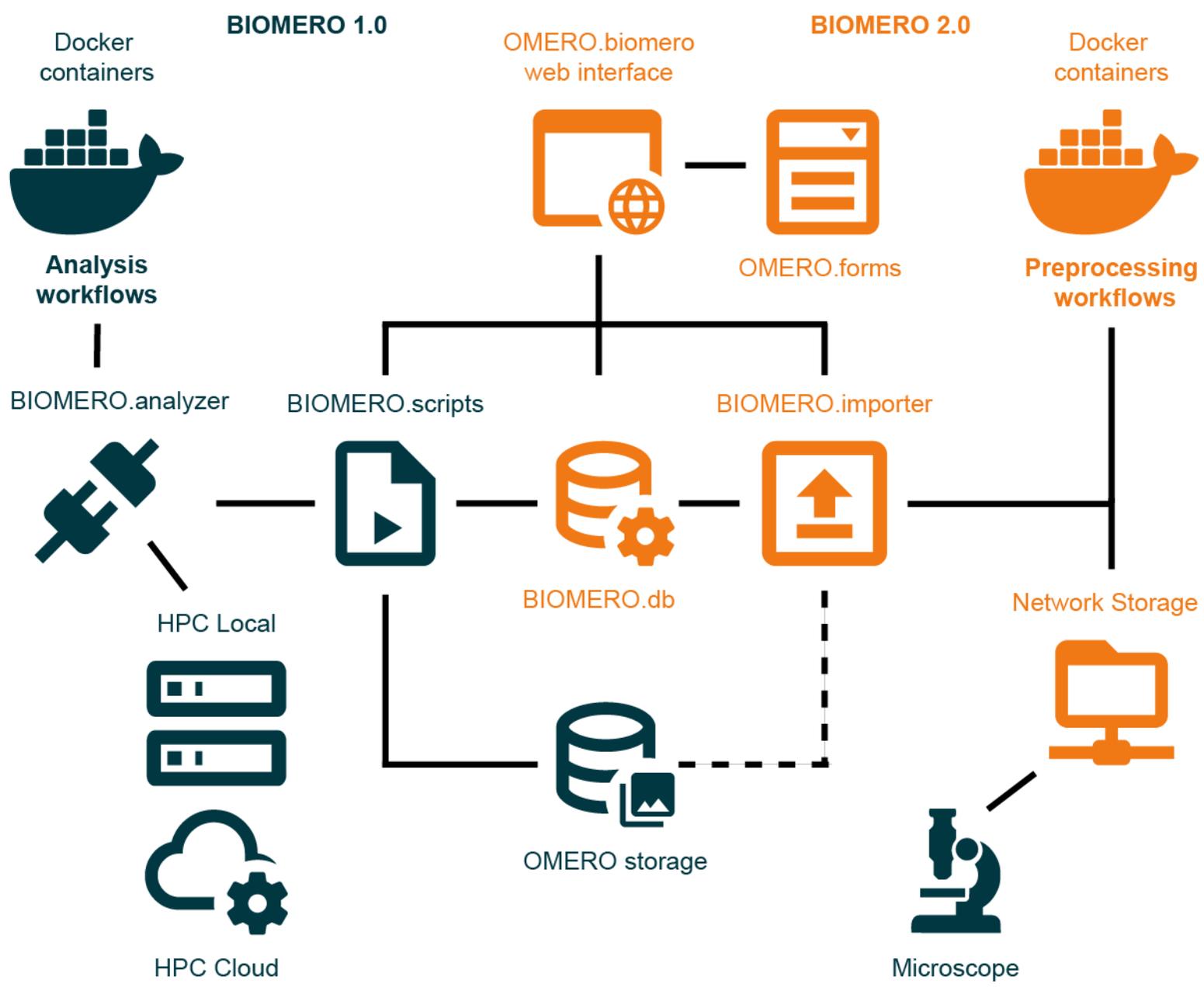

User selects input and destination in OMERO.biomero

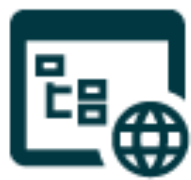

OMERO.biomero records order in BIOMERO.db

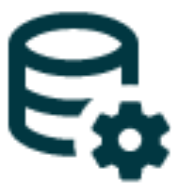

BIOMERO.importer retrieves new order

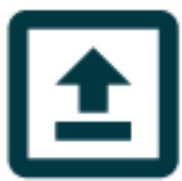

*BIOMERO.importer runs optional preprocessing*

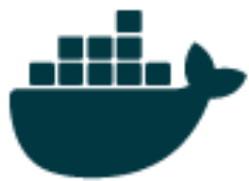

BIOMERO.importer performs in-place import into OMERO

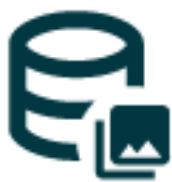

# Supplemental information

# BIOMERO 2.0: end-to-end FAIR infrastructure for bioimaging data import, analysis, and provenance

Torec T. Luik, Joost de Folter, Rodrigo Rosas-Bertolini, Eric A.J. Reits, Ron A. Hoebe, and Przemek M. Krawczyk

# Note S1. Preprocessing containers in BIOMERO.importer, related to the BIOMERO.importer enables automated preprocessing and in-place data import

We provide 2 example preprocessing containers for our needs: a Leica format converter[1] and a general OME converter[2].

Containers need to adhere to a number of required and optional input arguments about the input data, and to provide in provide output in JSON format[3].

The converter preprocessing container provides support for image formats not natively supported and converts these into OME-TIFF or OME-Zarr. The converter pipeline is structured in a modular way, containing source readers and writers for different formats. Once the source format is detected, a corresponding source reader is created, and a writer is created that matches the desired output format. The converter then uses the source reader and writer to convert either plates or single images. Additional source readers can easily be added by reusing functions that return metadata and image data in a prescribed way.

For use in the OMERO.biomero web plugin, conversion containers need to be configured in biomero-config.json[4]

# Figure S1. Workflow of the BIOMERO.importer for in-place import and preprocessing, related to the BIOMERO.importer enables automated preprocessing and in-place data import

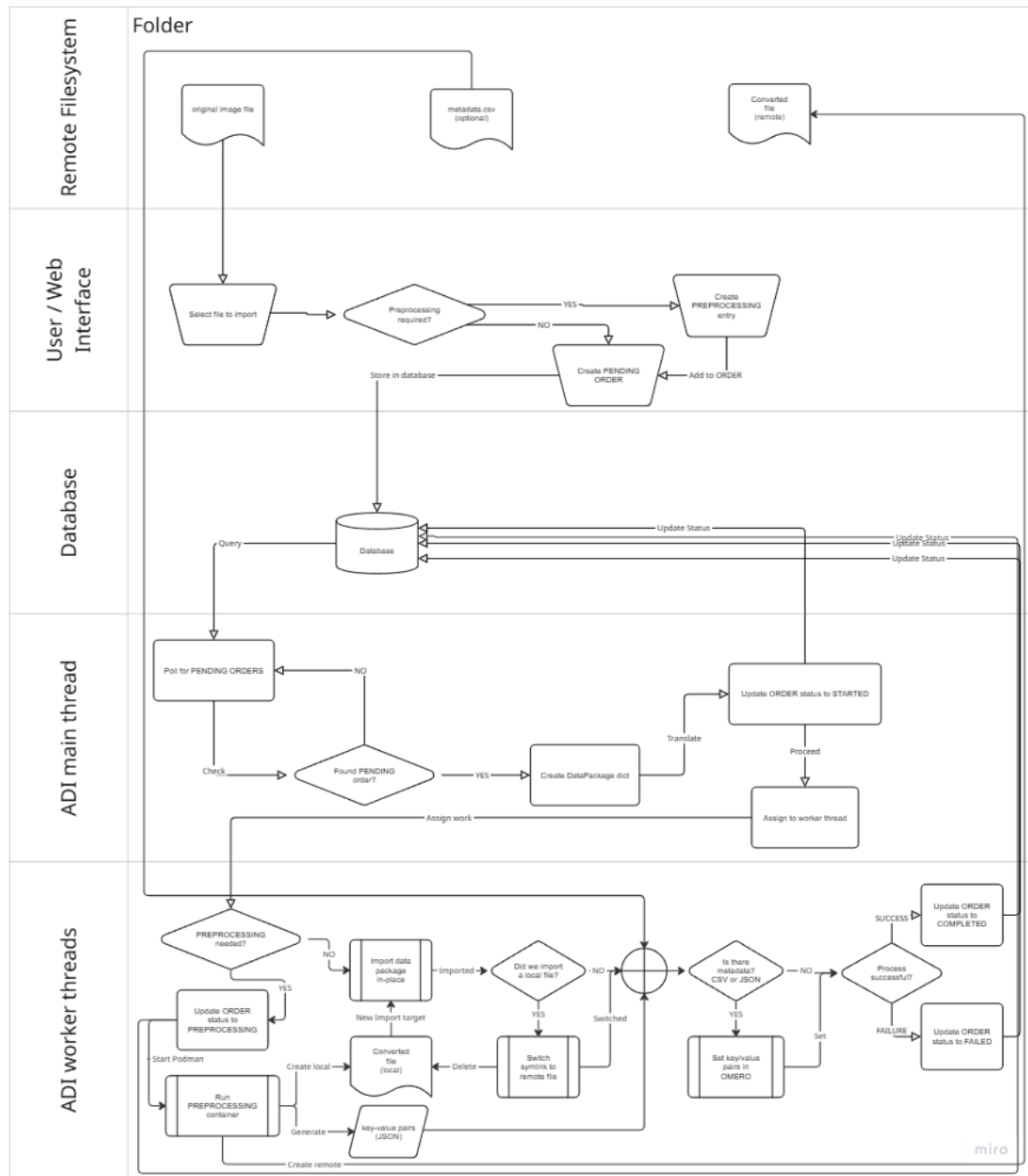

Figure S1 – Workflow of the BIOMERO.importer for in-place import and preprocessing. The flowchart illustrates the full data and decision flow leading to an in-place import. An image file located on the Remote Filesystem is selected by the User through the OMERO.biomero web interface. The interface creates a PENDING ORDER in the BIOMERO.db via the OMERO.biomero API,

including optional preprocessing settings for specific file formats. The BIOMERO.importer service continuously polls the database for such pending orders; once detected, an order is transformed internally into a DataPackage and updated externally to status STARTED. Parallel worker threads handle multiple imports concurrently. The assigned worker checks whether PREPROCESSING is required. If so, the status changes to PREPROCESSING, and a Podman container is launched on the importer host as defined in the database. This container may generate a converted file (stored locally and on the Remote Filesystem in a subfolder) and optionally produce metadata key–value pairs in JSON format. If a converted file is generated, it becomes the new target file in the DataPackage. Subsequently, the importer performs an in-place import using the OMERO.cli, linking either the original or converted file into OMERO. OMERO creates database entries and a symbolic link to the imported file. If this link points to a locally converted file, it is redirected to the remote converted file, and the local copy is removed. After the import, the importer attaches available metadata to the OMERO object—both JSON key–value pairs and, when present, a metadata CSV file located alongside the original image. Finally, the order in the database is updated to COMPLETED or FAILED, depending on the process outcome.

# Figure S2. Web interface for importer in OMERO.biomero, related to User Interface enables real accessibility

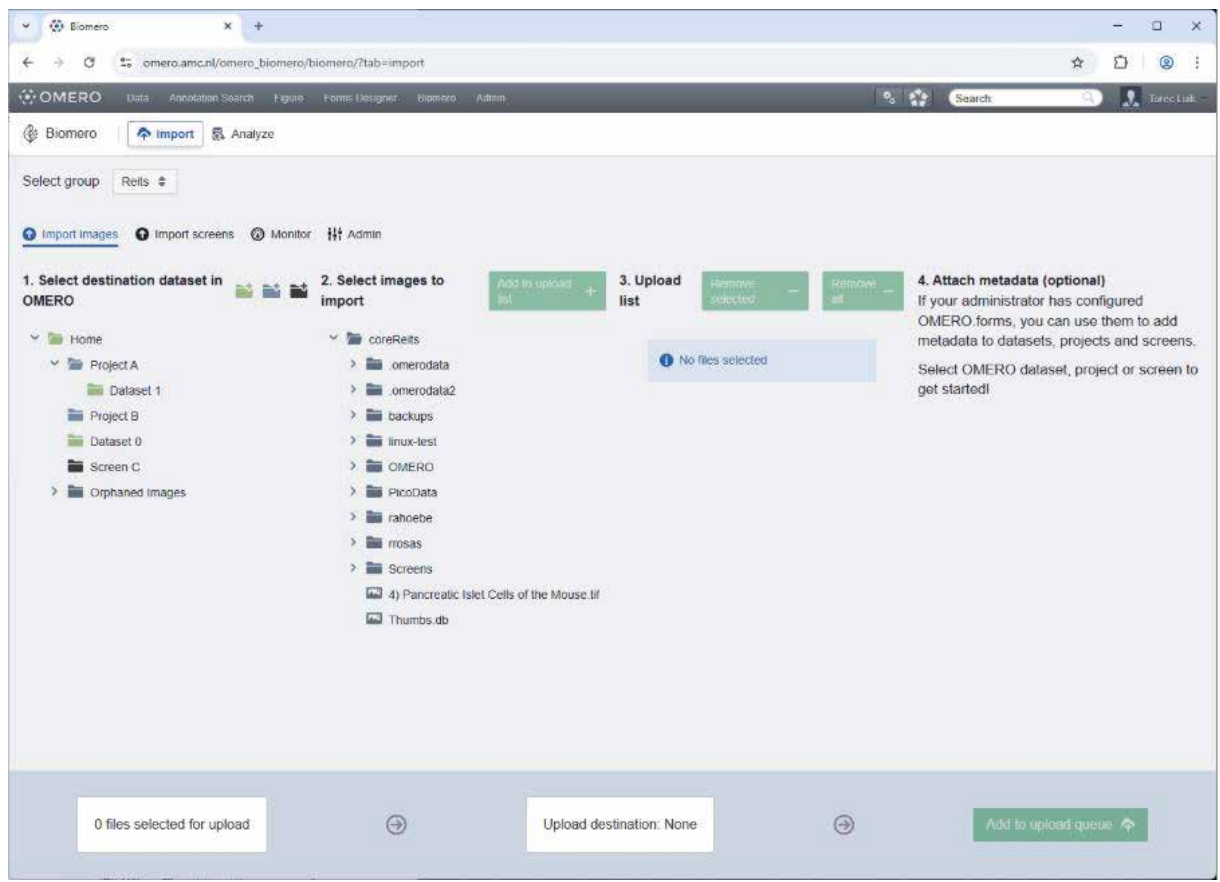

Figure S2 – User interface of OMERO.biomero, importer tab, import images tab, showing user's OMERO data structure and accessible remote data folder to import from. No selections or lists yet.

# Figure S3. Data selection for importer in OMERO.biomero, related to User Interface enables real accessibility

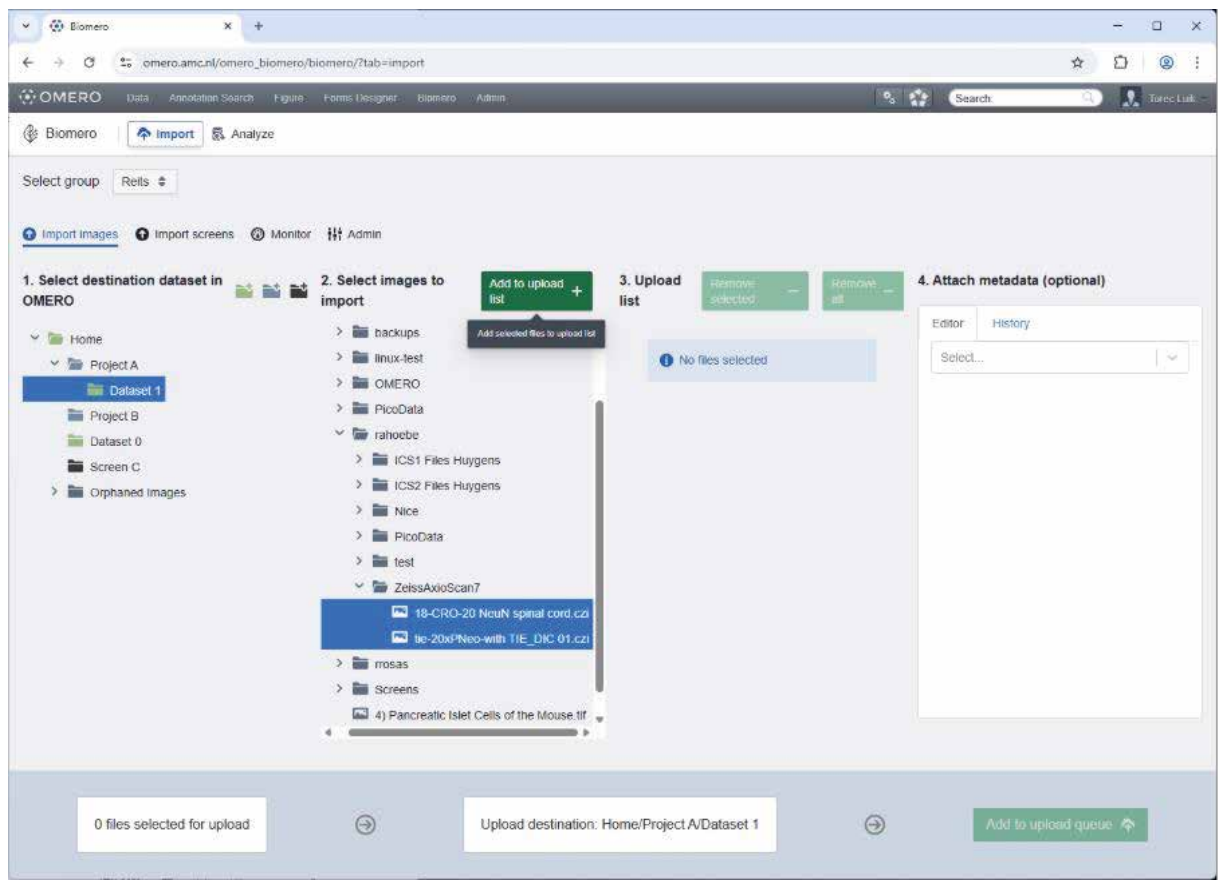

Figure S3 – Select destination dataset in OMERO, image on remote disk, ready to Add to Upload List

# Figure S4. OMERO.forms in OMERO.biomero, related to User Interface enables real accessibility

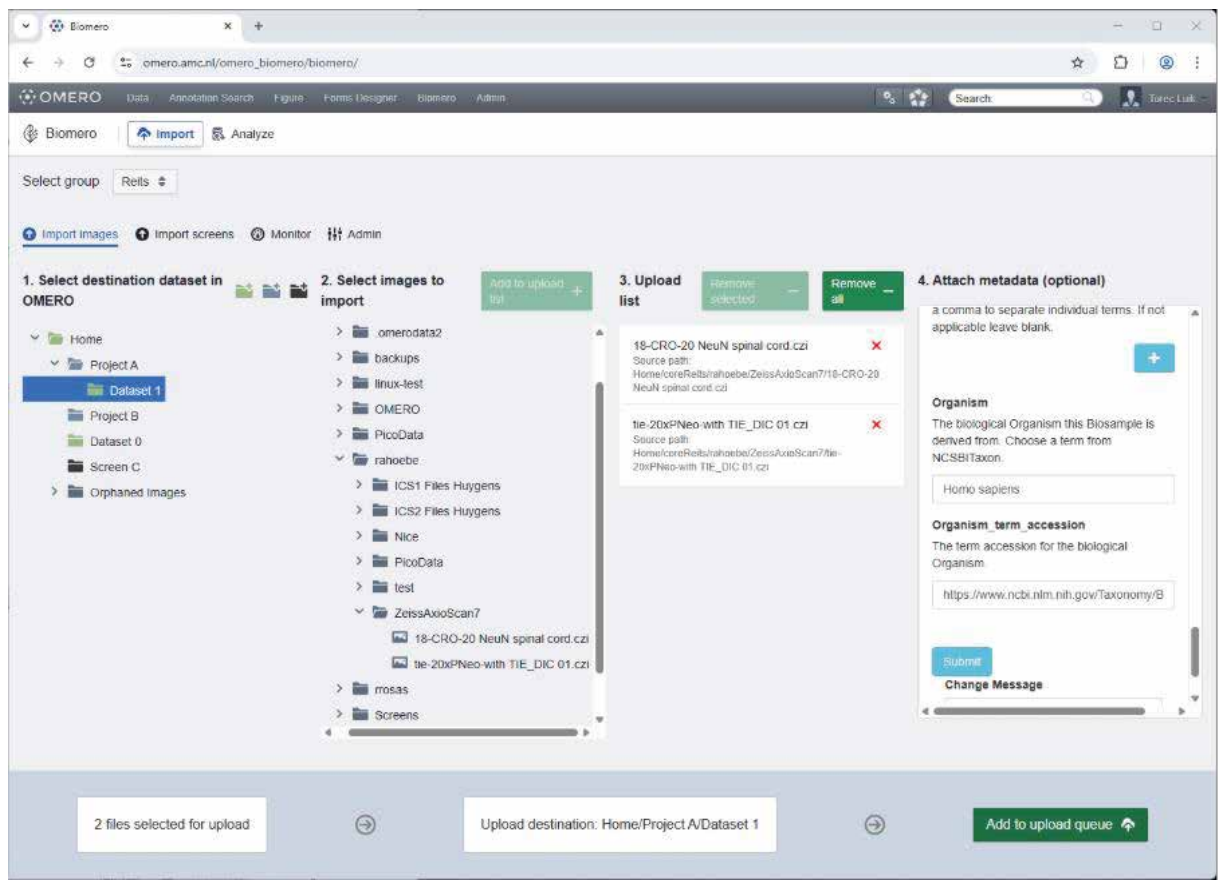

Figure S4 – Selection is added to the Upload list. A metadata form is filled in for the selected Dataset. User is ready to Add to upload queue. User can click Add to upload queue at the bottom right to submit the import order and can continue filling in the metadata form and Submit it separately.

# Figure S5. Import screens in OMERO.biomero, related to User Interface enables real accessibility

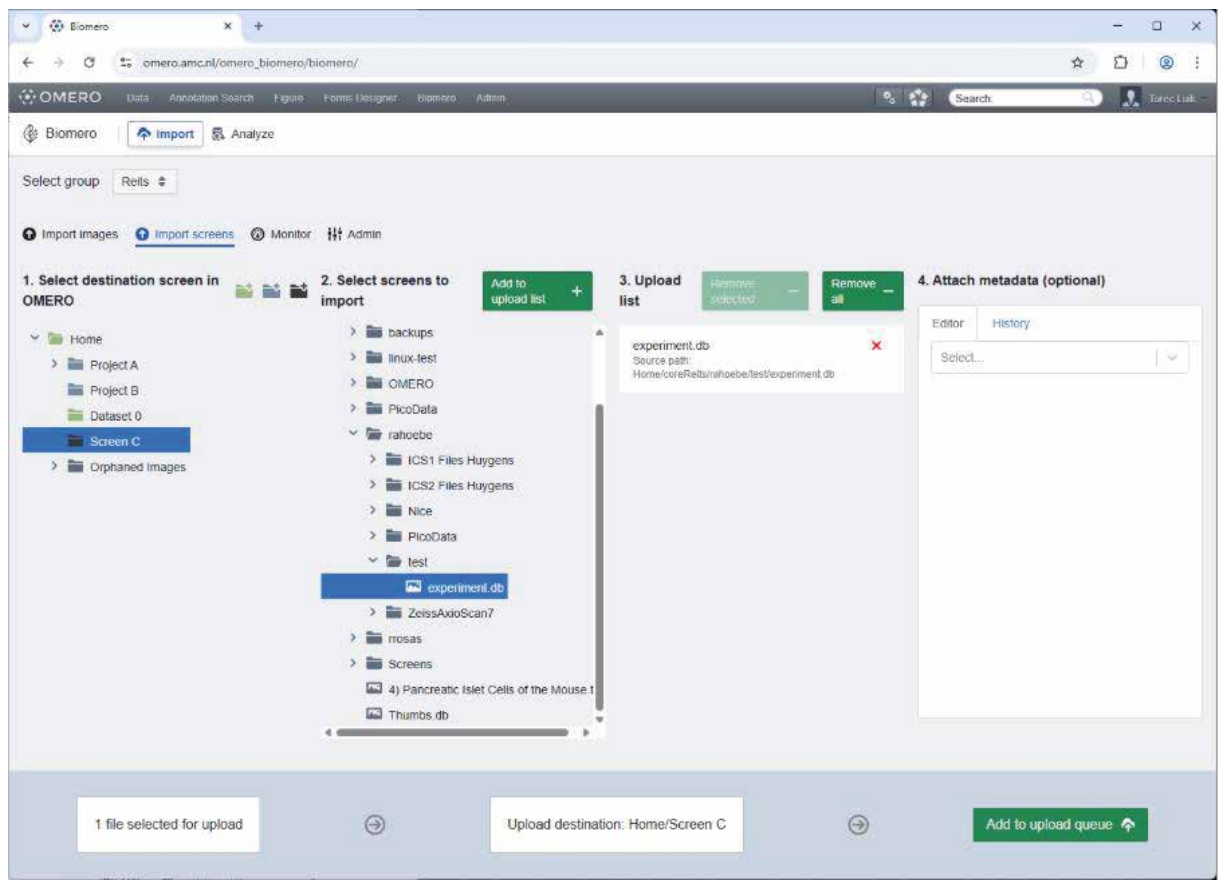

Figure S5 – Second tab to Import screens. Screen is selected as destination in OMERO. A plate has been selected to import and added to the upload list. Metadata forms could be filled in and the screen can be added to the upload queue.

# Figure S6. Monitor tab for importer in OMERO.biomero, related to User Interface enables real accessibility

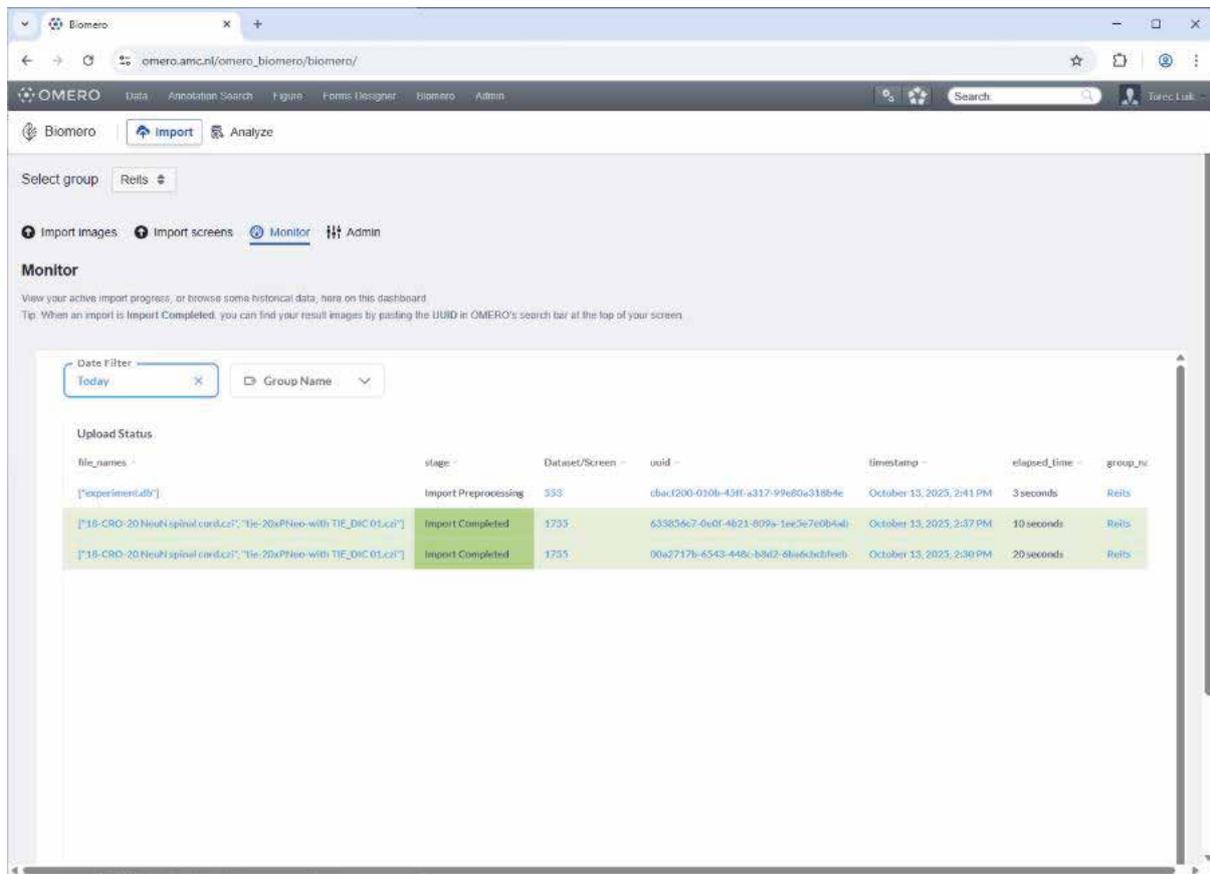

Figure S6 – Third tab to Monitor the progress of imports shows our Metabase dashboard, filtered to Today. We see the stage Import Preprocessing for the screen "experiment.db", with UUID starting with cbacf200. We see the stage Import Completed for 2 similar CZI file orders for Dataset 1755, each with a unique UUID and timestamp.

# Figure S7. Search UUID for importer in OMERO, related to User Interface enables real accessibility

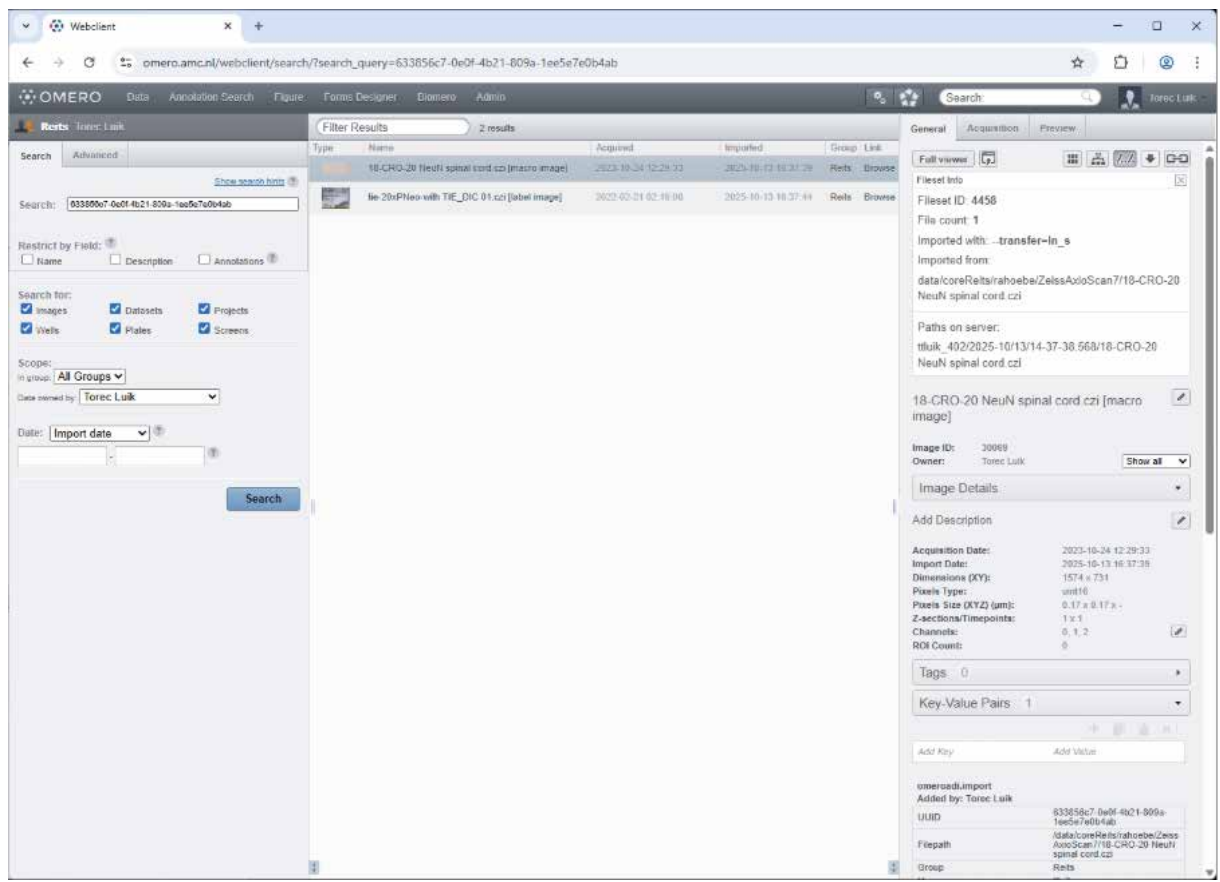

Figure S7 – We can click on the UUID to search OMERO for this UUID value as shown in the screenshot. This finds the 2 different CZI file(set)s that were part of the import order. Metadata on the top right side shows that this data has been imported with "—transfer=ln_s", which indicates the in-place import. Metadata on the bottom right side shows Key-Value Pairs including the UUID that we searched for.

# Figure S8. Metadata for importer in OMERO, related to User Interface enables real accessibility

Figure S8 – More in-depth view of the metadata via the Key-Value Pairs that were added as part of this import. We can see the omeroadi.import namespace providing the basic information of the import order and who executed it. We also see the metadata added to the parent dataset from the OMERO.form the user filled in; in this example detailing the Organism as Homo sapiens with a taxonomy link.

# Figure S9. Admin tab for importer in OMERO.biomero, related to User Interface enables real accessibility

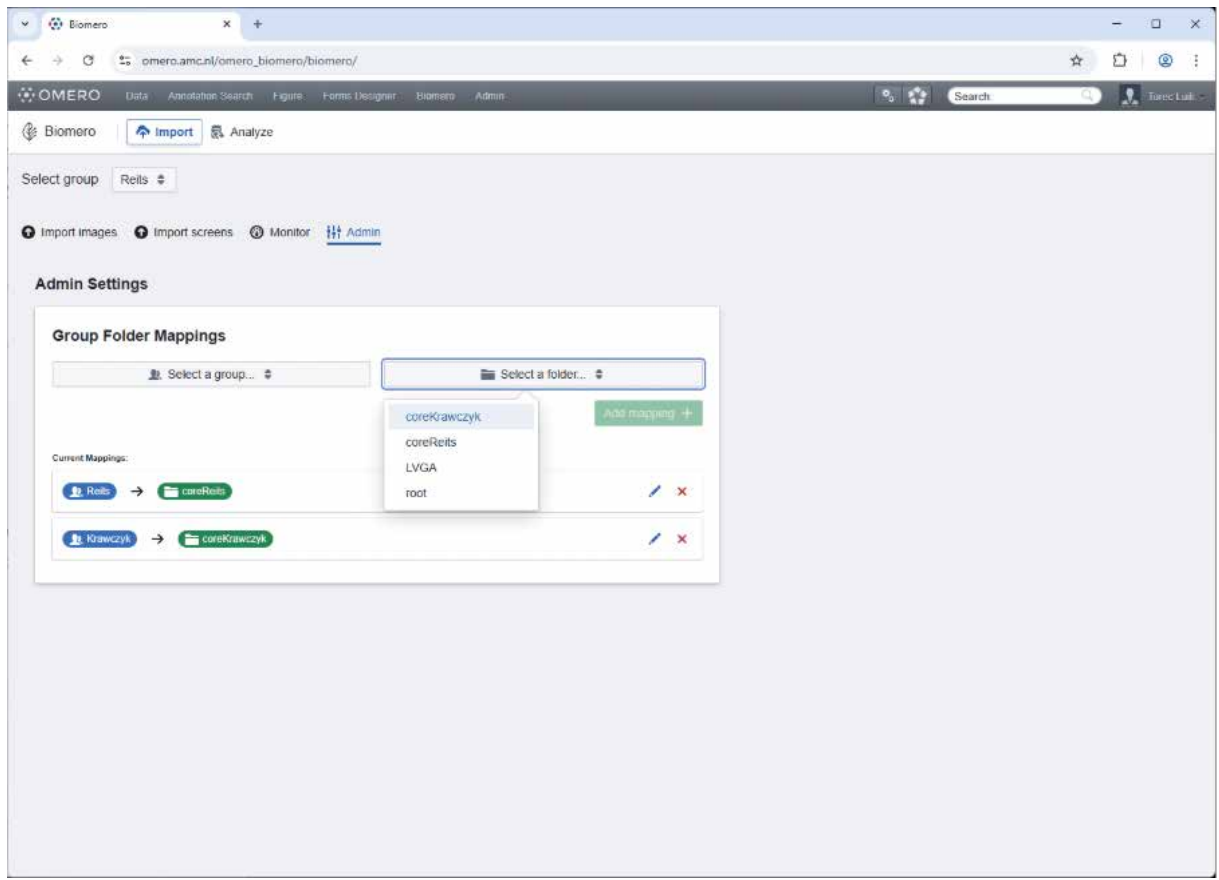

Figure S9 – The Admin tab of the Import section. We can see 2 existing mappings of OMERO group to a subfolder: group Reits to folder coreReits and group Krawczyk to folder coreKrawczyk. We can edit or delete these existing mappings with respectively the pen and x icons. There are more folders we haven't assigned to an OMERO group yet and vice versa; we can create a new mapping by selecting a combination and clicking the Add mapping button.

# Figure S10. Run tab for analyzer in OMERO.biomero, related to User Interface enables real accessibility

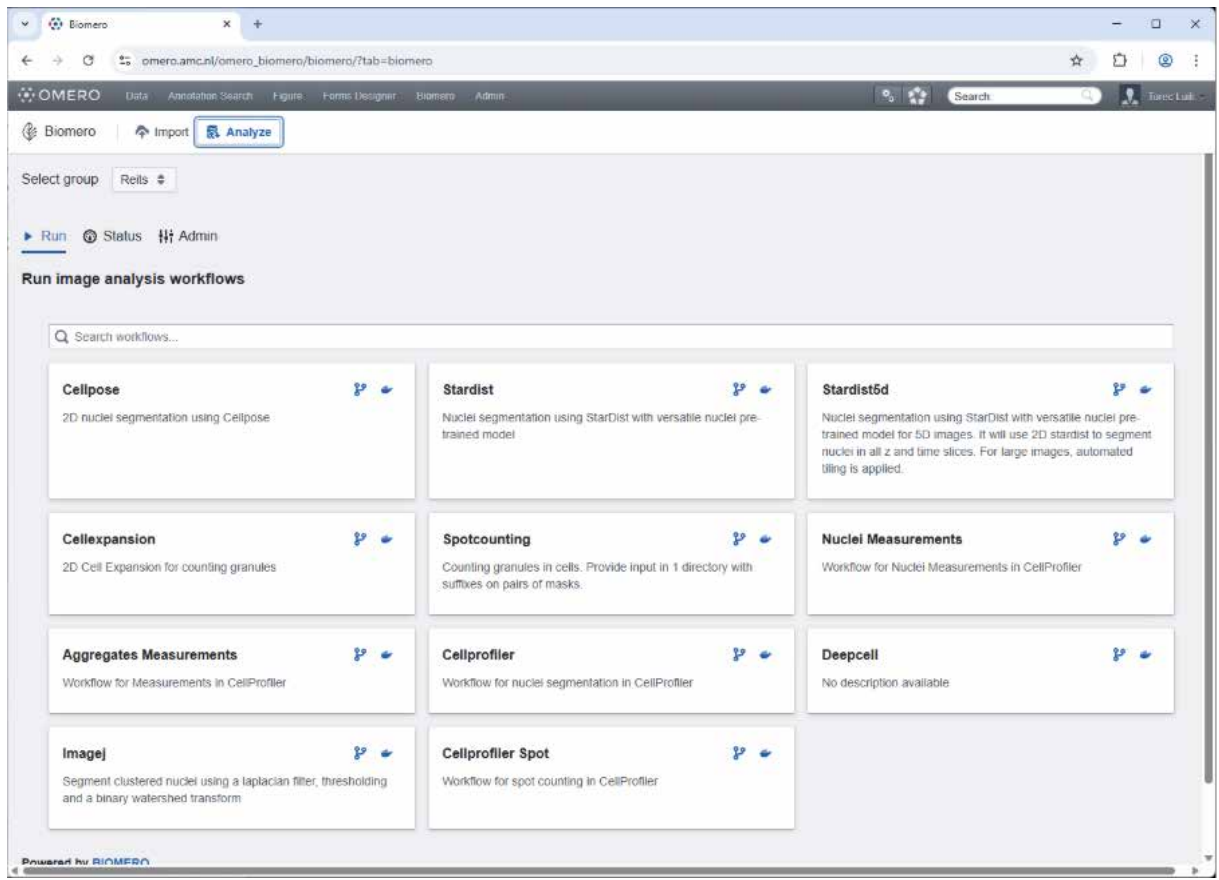

Figure S10 – The Run tab of the Analyze section. We can see the available selection of workflows cards that the user can select, including Cellpose and Stardist.

# Figure S11. Filter workflows in OMERO.biomero, related to User Interface enables real accessibility

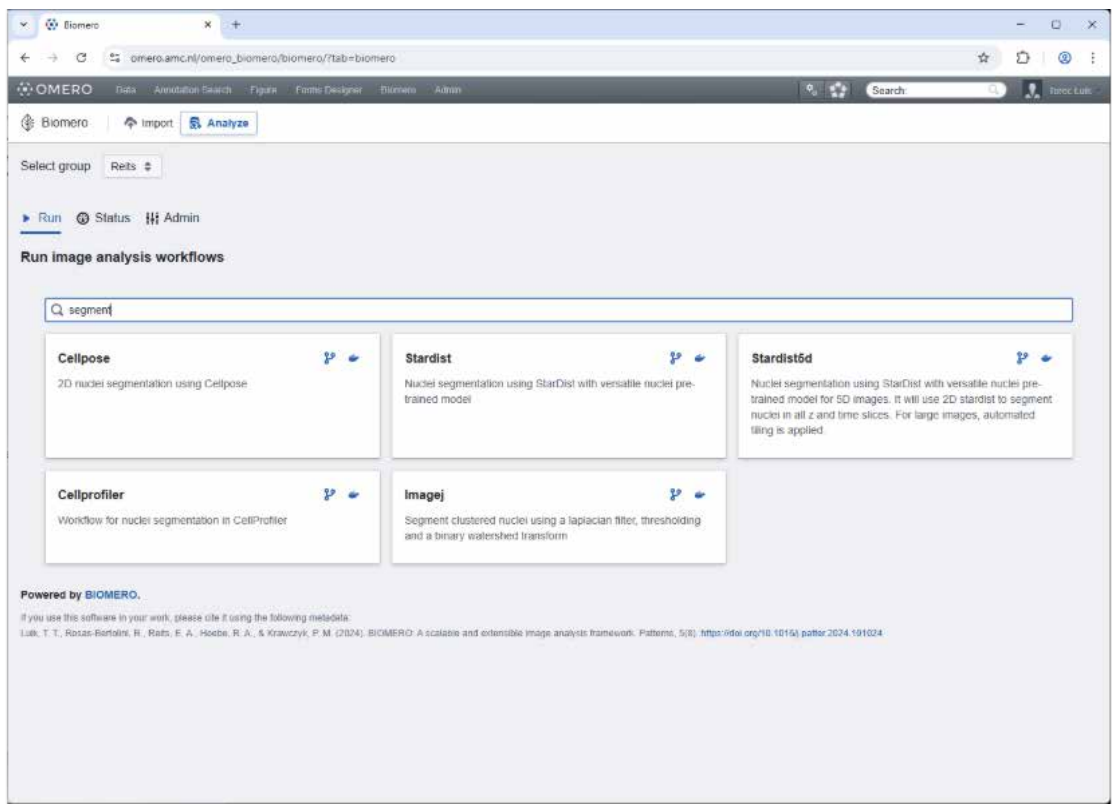

Figure S11 – Filtered workflows. We can see that the available workflow cards are filtered by the search term the user has entered. Only segmentation options are shown.

# Figure S12. Workflow input selection in OMERO.biomero, related to User Interface enables real accessibility

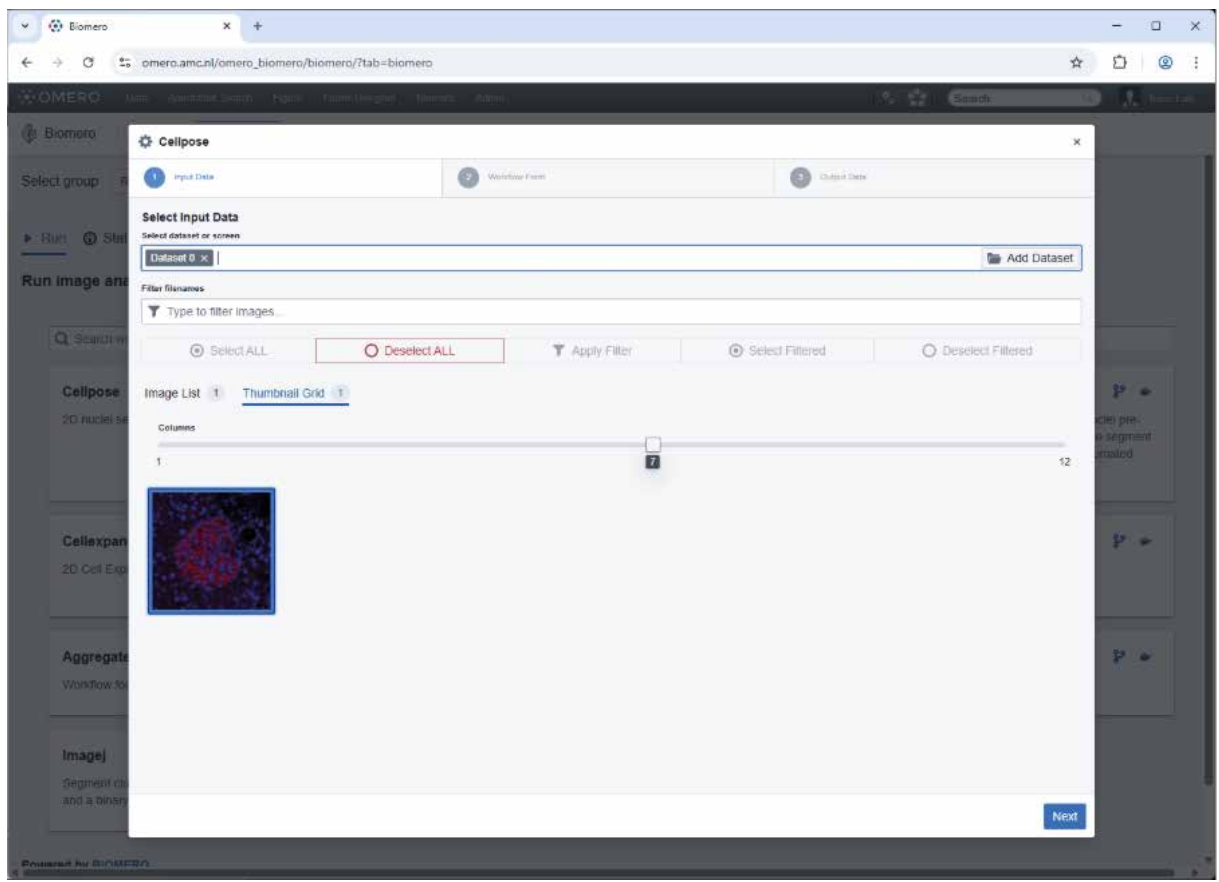

Figure S12 – Workflow dialog box. The Cellpose workflow has been selected, which opens the dialog box showing 3 steps to be taken: Input Data, Workflow Form and Output Data. For the Input data the user must select an OMERO dataset or screen. Dataset has been selected. Thumbnail is shown of the images it contains. All images in the dataset are selected by default but can now be unselected. User continues with the Next button.

# Figure S13. Workflow parameter selection in OMERO.biomero, related to User Interface enables real accessibility

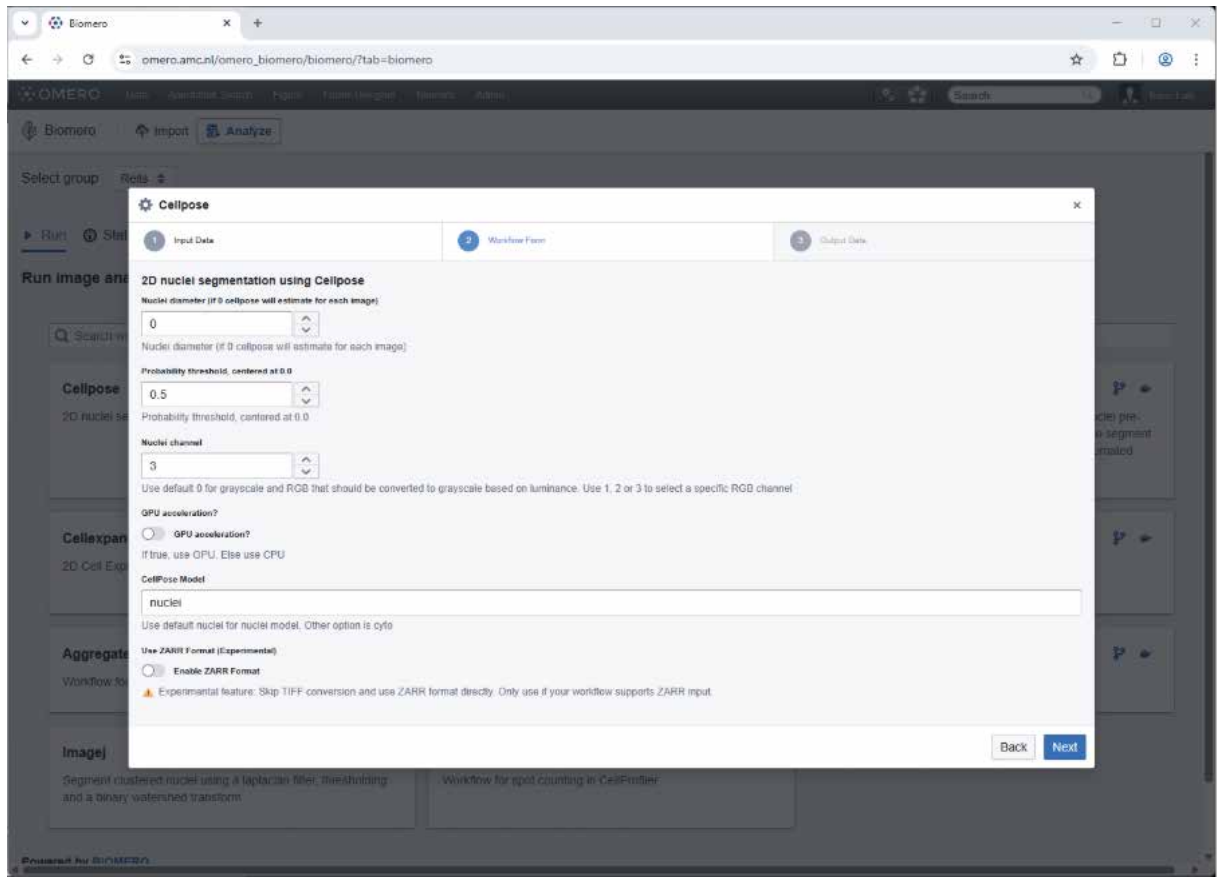

Figure S13 – Workflow Form step of the Dialog. We see the Workflow Form as generated for the Cellpose BIAFLOW. User selected Nuclei channel to be 3. No GPU acceleration was selected and ZARR format was not enabled. Default nuclei model is used, with Nuclei diameter and Probability threshold also using default values.

# Figure S14. Workflow output selection in OMERO.biomero, related to User Interface enables real accessibility

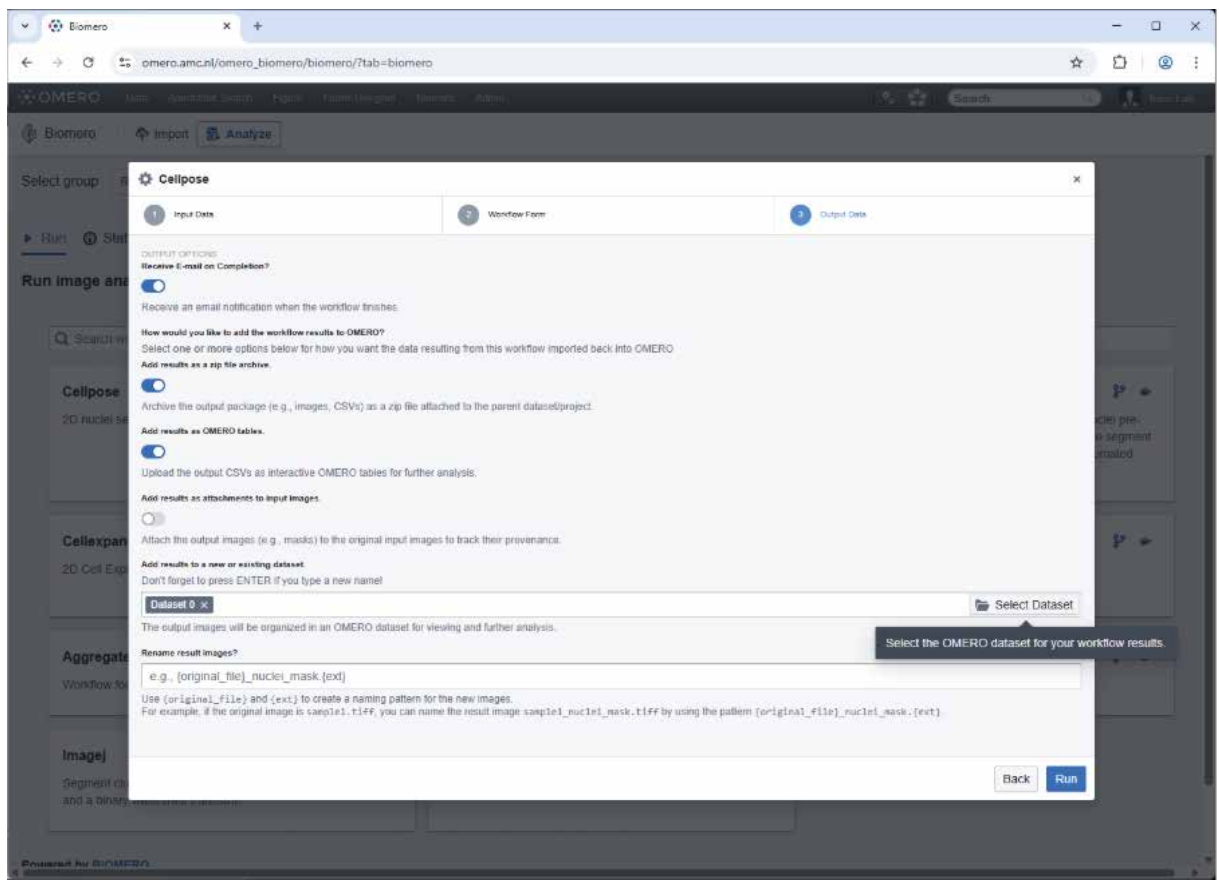

Figure S14 – Output Data step of the Dialog. We see the output options available to the user, with defaults enabled. Cellpose will generate a mask image of the nuclei and the input dataset (Dataset 0) has been selected to store this mask image in. The user also enabled receiving an e-mail from Slurm (if enabled on Slurm). Other output data options are enabled such as storing the results as a zip file archive and attaching results as OMERO tables. No rename option has been chosen, so the name will be whatever the workflow gives to the mask image. Workflow can be started by clicking the Run bottom in the bottom right.

# Figure S15. Status tab in OMERO.biomero, related to User Interface enables real accessibility

Figure S15 – Status tab of the Analyze section. After workflow has been submitted, status can be tracked on the Status tab. We can see the Metabase dashboard showing the Latest BIOMERO jobs for this user with no filters enabled yet. The top of the table shows the cellpose job that was submitted today, with current status JOB_COMPLETED. It has already updated a few times before and will update a few more times. Older results from earlier in the month are shown in the rows below that, with different Statuses including FAILED workflows. Colors indicate Status as well, for a quick overview.

# Figure S16. Filtered status tab in OMERO.biomero, related to User Interface enables real accessibility

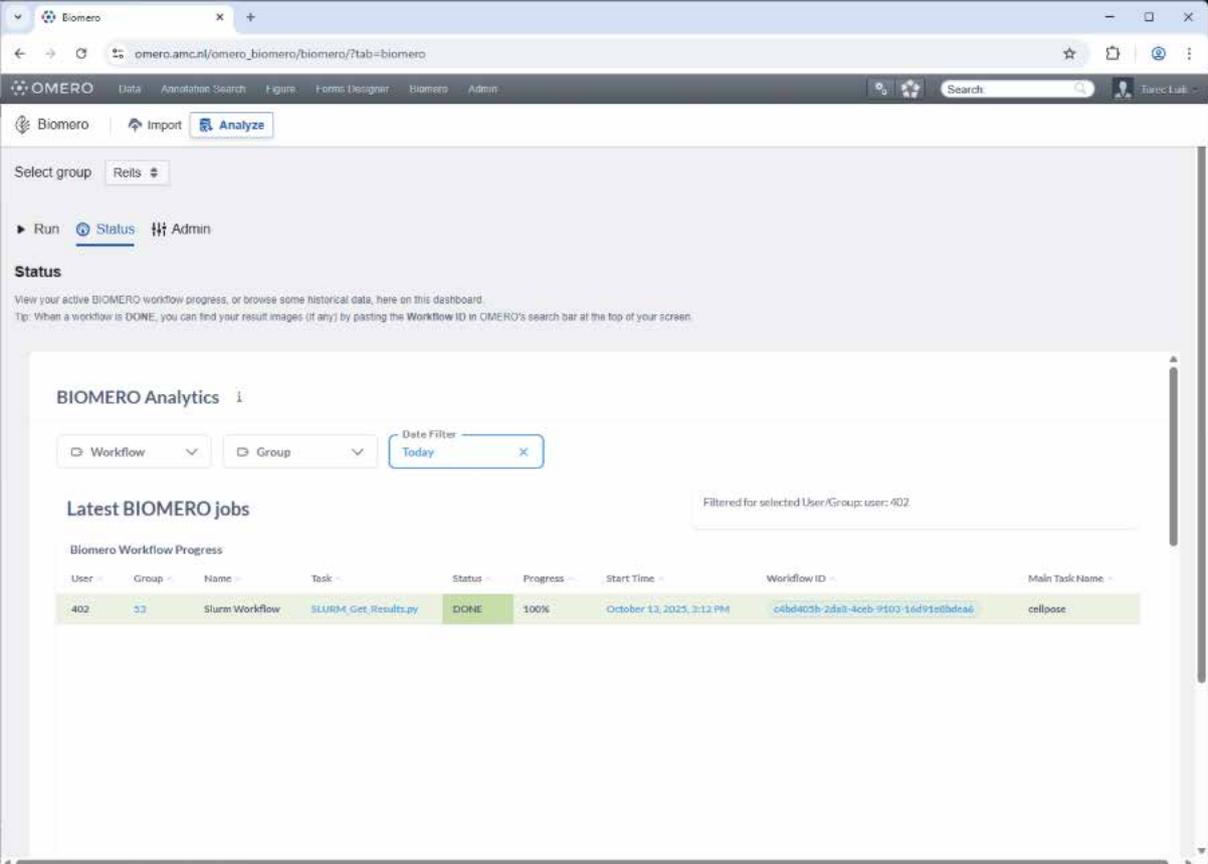

Figure S16 – Workflow is completed. We see the table filtered to Today. The Status of this workflow with UUID starting with c4bd405b has been updated to DONE and shown in green. As with the imports, also here we can redirect to the results by searching for this UUID (which also enabled by clicking on it).

# Figure S17. Search UUID for analyzer in OMERO, related to User Interface enables real accessibility

Figure S17 – By searching OMERO for this workflow UUID we can find the resulting mask image. We can see a subset of the Key-Value Pairs added by the BIOMERO workflow on this image: among others the UUID of the workflow that created it, the Version of the cellpose task and its chosen parameters such as Param_nuc_channel 3. Other metadata not shown includes the input image ID, the BIOMERO scripts used, the conversion container used, the specific Slurm sbatch command given and the cellpose container Slurm used and much more.

# Figure S18. View result images from analyzer in OMERO, related to User Interface enables real accessibility

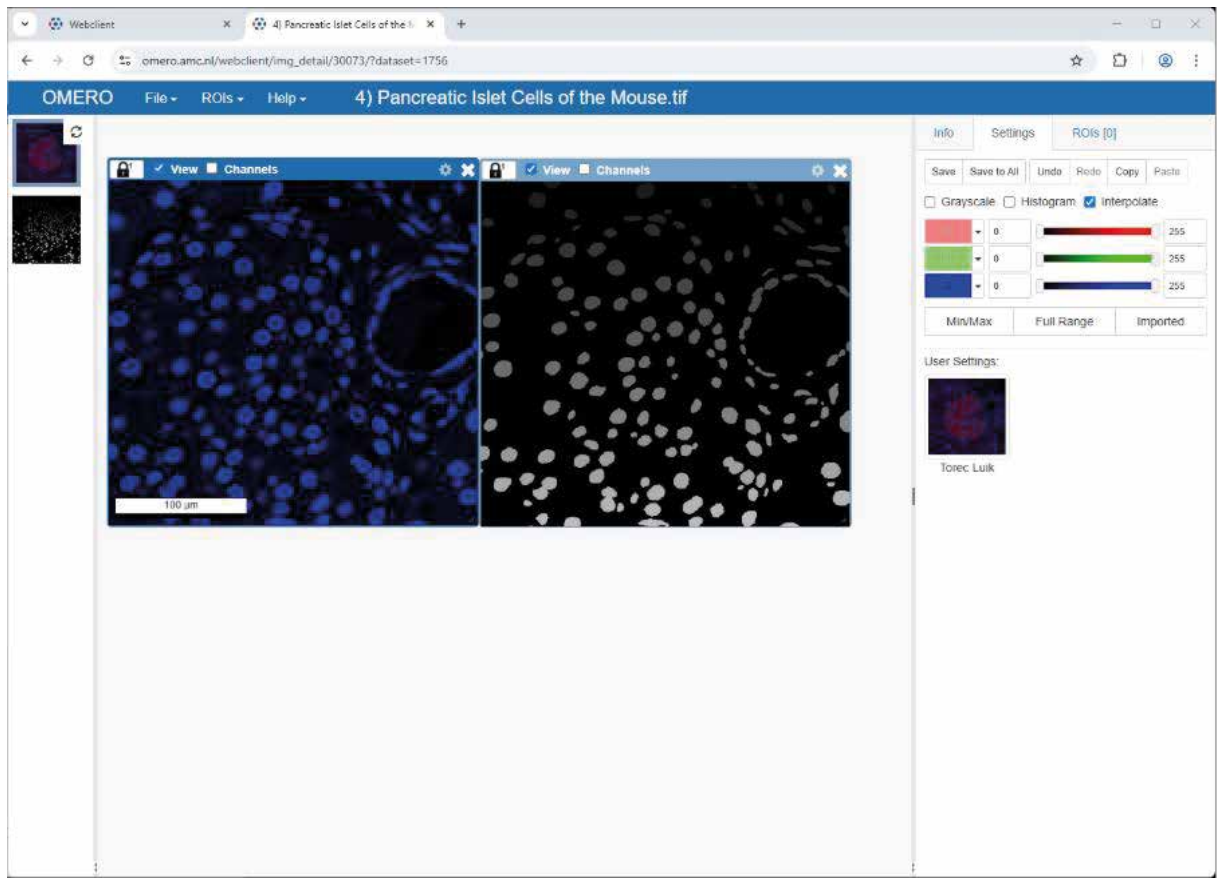

Figure S18 – OMERO viewer can be used to compare mask images to the original image to visually determine segmentation quality. We see a side-by-side interactive comparison of channel 3 versus the mask image in that same dataset.

# Figure S19. Admin tab for analyzer in OMERO.biomero, related to User Interface enables real accessibility

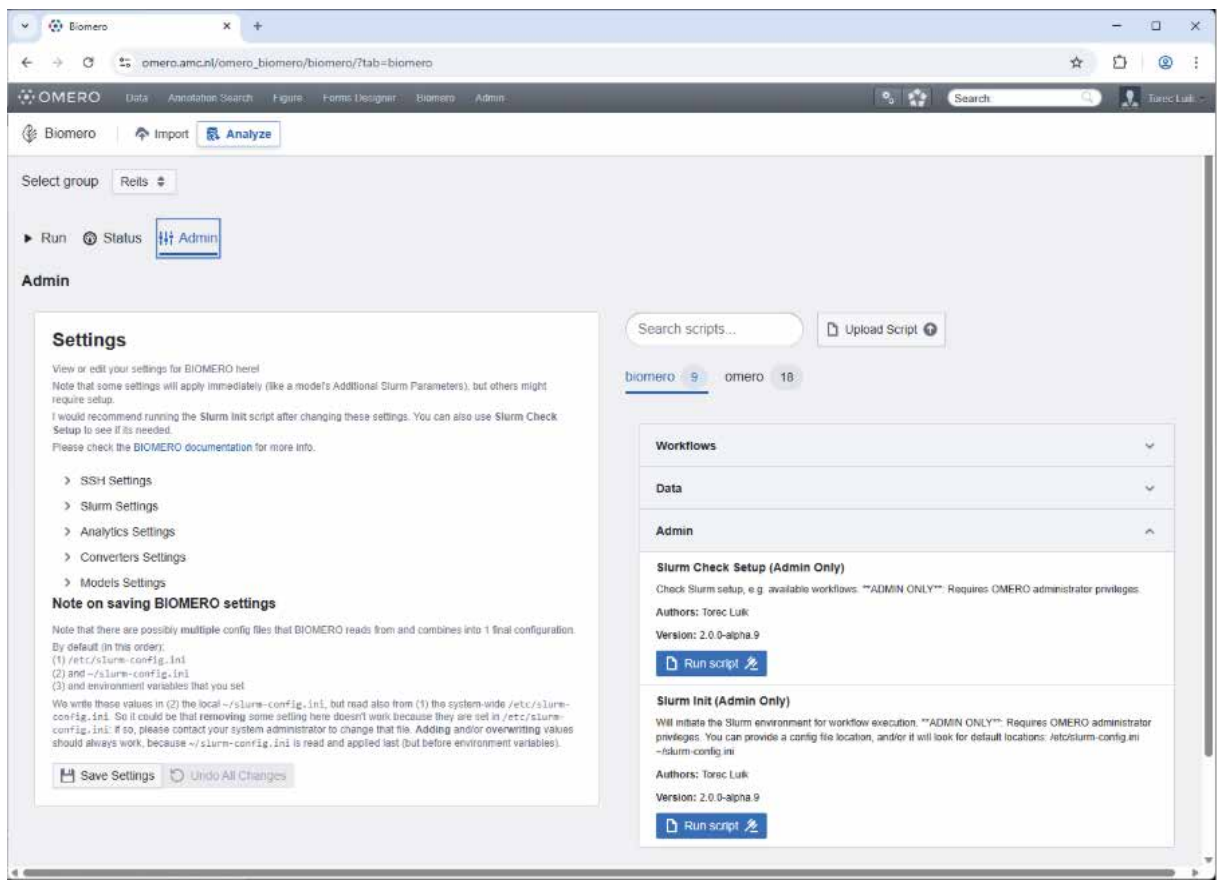

Figure S19 – Admin tab of the Analyze section. As an admin we can view this tab. On the left side we can see all the different sections of the BIOMERO settings file (slurm-config.ini). On the right side we can see a different filterable interface of OMERO.scripts, with the import Admin BIOMERO.scripts directly available to Run: Slurm Check Setup and Slurm init. After the admin click Save Settings on the left, it is likely these changes might requires Slurm changes: then they can immediately start the Slurm Init script on the right.

# Figure S20. Workflow model configuration in Admin tab in OMERO.biomero, related to User Interface enables real accessibility

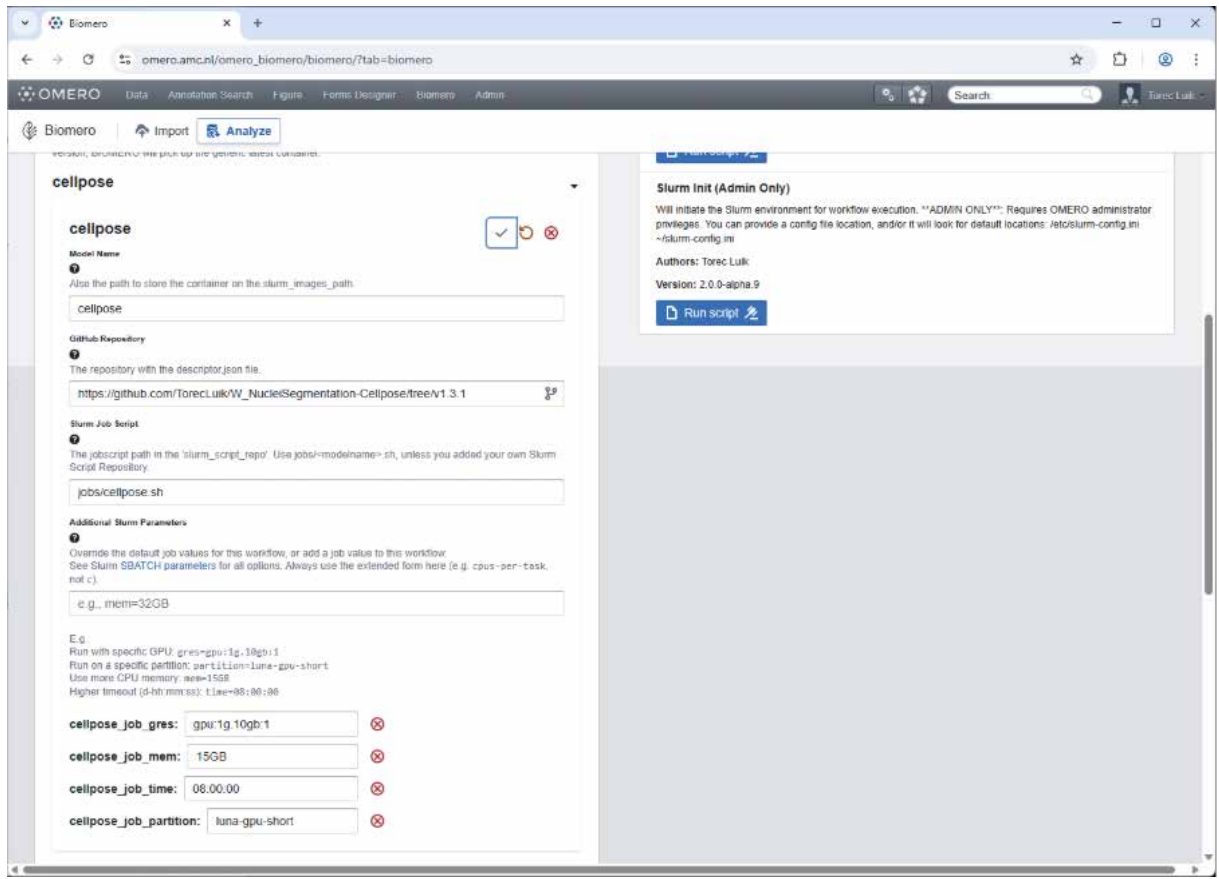

Figure S20 – Admin section on Model configuration for BIOMERO. We can see and edit the details of the cellpose workflow that we ran earlier. We can for example change the GitHub Repository to point to a new version, or we can change Slurm SBATCH parameters like the Slurm partition this workflow should run on. We could also add extra Slurm SBATCH parameters if needed, or remove existing ones.

# Figure S21. Add new workflow in Admin tab in OMERO.biomero, related to User Interface enables real accessibility

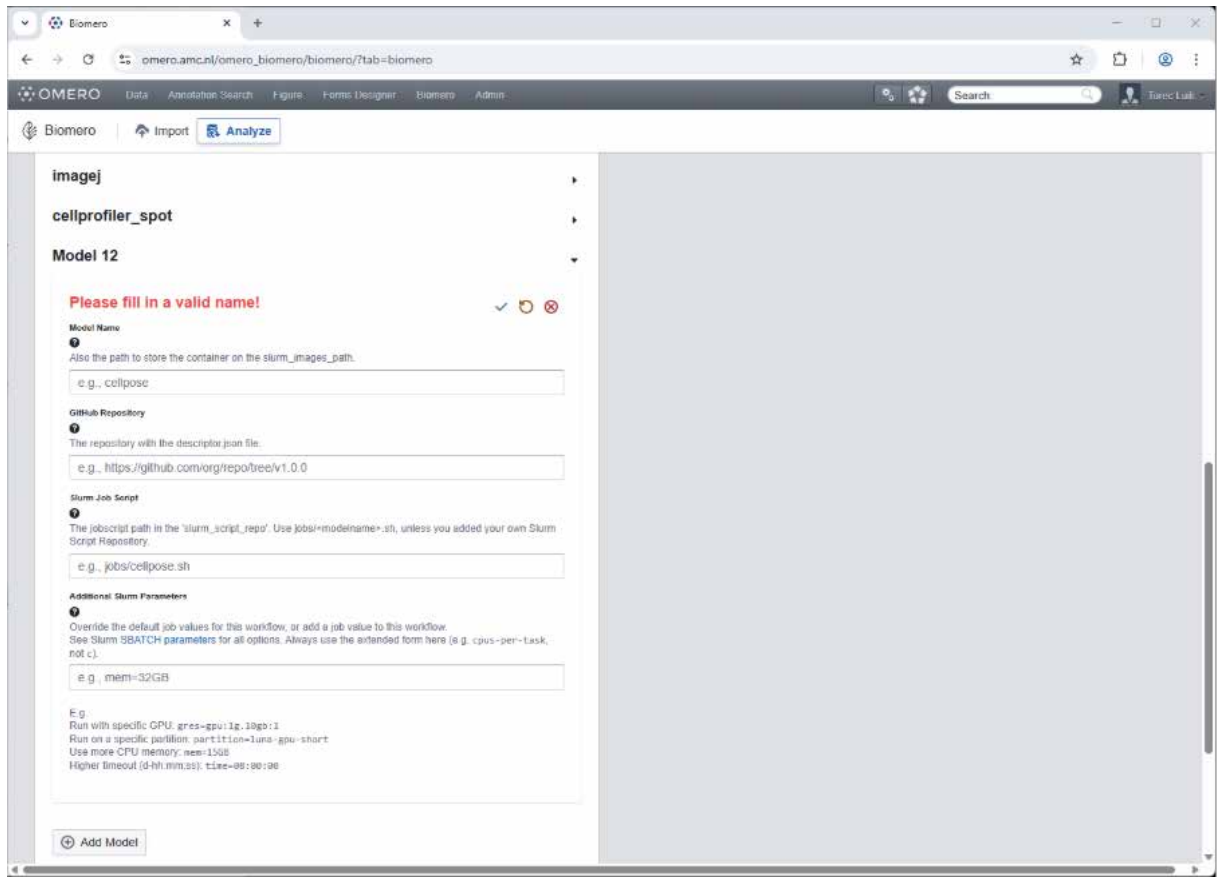

Figure S21 – Admin interface for adding a new Model. When the admin clicks the Add Model button, a new subsection is added for this new model with the 4 main fields ready to fill in: Model Name, GitHub Repository, Slurm Job Script and Additional Slurm Parameters. When the admin fills in the Model Name as shown, this automatically and immediately updates the section names and the Slurm Job Script field. When filling in the GitHub Repository, the warning shows that we have not provided a valid URL yet: it is expecting a GitHub URL pointing to a specific released version. When everything is filled in, the admin can Save Settings or Undo All Changes.

## Supplemental References

1. Cellular-Imaging-Amsterdam-UMC/ConvertLeica-Docker (2025). (Cellular Imaging, Amsterdam UMC).

2. NL-BioImaging/biomero-converter (2025). (NL-BioImaging AM).

3. General container requirements - biomero-converter https://nl-bioimaging.github.io/biomero-converter/container/.

4. OMERO.biomero Plugin Administration — NL-BIOMERO latest documentation https://cellular-imaging-amsterdam-umc.github.io/NL-BIOMERO/sysadmin/omero-biomero-admin.html#configuration-file-management.